\documentclass[letterpaper]{jpconf}
\usepackage{d0_style}
\usepackage{graphicx}

\begin{document}
\title{Physics with Single Photons plus Missing Energy Final States at \D0}

\author{Edgar Carrera for the \D0 Collaboration}

\address{Florida State University, Tallahassee, FL 32306, USA}

\begin{abstract}
Final state signatures of a single photon and missing  
transverse energy offer unique and powerful advantages 
in the search for new physics.  
This document presents the first 
observation of the $Z\gamma\to\nu\bar{\nu}\gamma$ 
process at the Tevatron Collider 
at $5.1$ standard deviations significance, as well as some of the
strongest limits on 
anomalous trilinear $ZZ\gamma$ and $Z\gamma\gamma$ 
couplings to date.  
Additionally, we present the latest \D0 results 
on a search for direct production of Kaluza Klein 
gravitons in association with single photons.

\end{abstract}

\section{Introduction}
In the standard model (SM), trilinear
gauge boson electroweak 
interactions involving photons and $Z$ bosons are
forbidden at the leading-order level. Since  
next-to-leading order corrections to $Z\gamma\gamma$ or $ZZ\gamma$ vertices
are small, an enhanced $Z\gamma$ production rate, particularly at higher
values
of the photon transverse energy ($E_{T}$),
would indicate anomalous values of the trilinear 
gauge couplings (ATGC) and therefore presence 
of physics beyond the 
SM~\cite{Gounaris:2002za,Choudhury:2000bw}.  These couplings
can be parametrized as eight complex numbers of the form, 
$h_{i}^{V}=h_{i0}^{V}/(1+\hat{s}/\Lambda^2)^{m}$~\cite{CT:baur:1993},
where $i=1,\ldots,4$ and $V=Z,\gamma$.  Here, 
the subscript ``$0$'' indicates
the low energy approximations of the couplings, $\hat{s}$ 
is the square of the parton center-of-mass energy, $\Lambda$ is 
the energy scale related to novel interactions responsible
for new physics, $m=3$ for 
$h_{1}^{V}$ and $h_{3}^{V}$, and $m=4$ for $h_{2}^{V}$ and 
$h_{4}^{V}$~\cite{CT:baur:1993}.  The choice of the form factor, 
with the given values
of $m$, guarantees the preservation of unitarity conditions at high energy.
We study, using $3.6\ {\rm fb}^{-1}$ of data, the
$Z\gamma$ production where the $Z$ boson decays 
to $\nu\bar{\nu}$.  This final state, whose collider signature
can be inferred by the presence of a single reconstructed photon
and missing transverse energy ($\met$),  is the most sensitive to 
ATGC, thanks to the large branching ratio (BR) 
of the $Z$ boson to the invisible channel.  This allows us to
set limits on the size of the real parts of the ATGC.

The single photon signature and missing energy 
can also be used to search for
the presence of
large extra spatial dimension (LED) in our universe.
In these scenarios~\cite{ArkaniHamed:1998rs}, SM particles are bound to the ordinary
3-dimensional space, while the gravitational field exists as quantized
towers of Kaluza-Klein (KK) modes (known as KK gravitons, $G_{KK}$)  
in the extra space created by $n$ large extra dimensions of size $R$.
The large unexplained difference 
between the effective Planck mass scale in the
$4$-dimensional space-time ($M_{Pl}\sim 10^{19}\ {\rm GeV/c^2}$)
and the electroweak scale ($\sim 10^3\ {\rm GeV/c^2} $), can be
solved  since the fundamental
Planck scale in the $(4+n)$-dimensional 
space-time ($M_{D}$), which could
be of the order of the electroweak scale,
is concealed by the large size of the extra volume:
$M_{Pl}^2 = 8\pi M_{D}^{n+2}R^{n}$.
In this analysis we set limits to $M_{D}$, using 
$2.7\ {\rm fb}^{-1}$ of data, by studying the 
$Z\gamma$-analogous process
$q\bar{q}\to G_{KK}$ where the $G_{KK}$ is not detected.  These results
constitute 
an update to Ref.~\cite{Abazov:2008kp}.

\section{Event Selection}

The data used in both analyses are separated in two
different datasets with different luminosity profiles,
which result in different efficiencies and backgrounds.
They were collected with
the \D0 detector~\cite{Abazov:2005pn}, which consists of a central tracking system
housed within a $2\ {\rm T}$ superconducting solenoid magnet,
a preshower system,
a liquid-argon calorimeter with fine-segmented electromagnetic 
(EM) and hadronic sections and a muon spectrometer with
its own $1.8\ {\rm T}$ iron toroidal magnet.  Events are
required to pass trigger sets that demand at least one energy
cluster in the EM section of the calorimeter with 
$E_{T}>20\ {\rm GeV}$.  These trigger requirements are 
fully efficient for photons with $E_{T}>90\ {\rm GeV}$.

Photons are identified in the calorimeter as narrow clusters
with at least $95\%$ of their energy in the EM section.  These
clusters are required to be isolated in the calorimeter, 
{\it i.e.}, the isolation variable,
${\cal{I}} =(E^{\rm tot}_{0.4}-E^{\rm em}_{0.2})/E^{\rm em}_{0.2}$, 
is required to be less than $0.07$.  In this equation,
$E^{\rm tot}_{0.4}$ denotes the total energy deposited
in the calorimeter in a cone of radius
${\cal{R}}=\sqrt{(\Delta\eta)^{2}+(\Delta\phi)^{2}} = 0.4$~\cite{defeta} and
$E^{\rm em}_{0.2}$ is the EM energy in a cone of radius
${\cal{R}}=0.2$.    
The track isolation variable, defined as the scalar sum
of the transverse momenta of
all tracks that originate from 
the interaction vertex in an hollow cone
of $0.05<{\cal{R}}<0.4$ around 
the cluster, is required to be less than $2\ {\rm GeV}$.
The only EM clusters considered  are 
the ones reconstructed within the central
region of the detector 
($|\eta|<1.1$)
and consistent with the longitudinal and transverse 
shower shape profiles of a photon. Additionally,
we require that there is no reconstructed track associated with
the cluster nor a significant density of hits in the tracking
system in agreement with that of 
a charged track.  The EM cluster is
required to have an associated 
energy cluster in the central preshower
system (CPS).

We select the photon sample by requiring events with a 
a photon candidate of 
$E_{T}>90\ {\rm GeV}$ and $\met>70$, which effectively
suppresses the multijet background.  The $\met$ is computed from
calorimeter cells and corrected from EM an jet energy scales.
Jets with $E_{T}>15\ {\rm GeV}$ are rejected in order to
minimize large $\met$ due to jet energy mismeasurement.
Events with muons, cosmic ray muons and energetic additional
EM objects or tracks are rejected as well. 

\section{Analysis}

In order to reduce the large background from cosmic ray muons
and beam halo particles 
(together referred as 
non-collision background) that deposit energy in the calorimeter
and mimic the signal signature, we require at least one
interaction vertex consistent with the direction of the photon
as given by the ``pointing'' algorithm.  
Assuming that EM showers are originated
by photons, the EM pointing algorithm predicts the distance
of closest approach to the $z$ axis along the beam line and
the $z$ position of the interaction vertex in the event,
independently of the tracker information and based solely
on the calorimeter and the CPS.
The remaining non-collision background and the backgrounds that
arise from jet misidentification in $W/Z$ events, are estimated
by fitting the photon sample DCA distribution, 
as it is explained in detail in Ref.~\cite{Abazov:2008kp},
to a linear sum of three DCA templates: a signal-like template,
a non-collision template and a misidentified jets template.  The
procedure estimates the fractional contribution of these
components.

The number of 
background events from the electroweak process $W\to e\nu$,
where the electron is misidentified as a photon, is
estimated by selecting an electron sample with the same
kinematical requirements as the photon sample, and
scaling  the final number of events by the measured
rate of electron-photon misidentification.  
The $Z\gamma\to\nu\bar{\nu}\gamma$ (for the LED
analysis only) and the $W\gamma$
background, where the $W$ boson decays leptonically and
the lepton is not detected, 
are estimated using samples of Monte Carlo
(MC) events, which
are generated using {\sc pythia}~\cite{Sjostrand:2000wi}, put through
a {\sc geant}-based~\cite{geant} \D0 detector simulation program
and reconstructed with
the same software as used for data.  Differences between
data and simulation are corrected by applying scale factors. 
A summary of backgrounds, for both analyses, is shown in
Table~\ref{tab:bkg}.

\begin{table}[!Hhtb]
\small
\begin{center}
\caption{\label{tab:bkg} \small Summary of background estimates, 
and the number of observed and SM predicted events. The 
numbers represent a combination of two separate data sets with different profiles of the 
instantaneous luminosity.}
\begin{tabular}{|c|c|c|}
\hline
  Process            &  ATGC Analysis $3.6\ {\rm fb}^{-1}$ & LED Analysis $2.7\ {\rm fb}^{-1}$\\ \hline
$Z\gamma\to\nu\bar{\nu}\gamma$ & --& $29.5\pm 2.5$\\
$W \to e\nu$      & $9.67 \pm 0.56$ & $8.5 \pm 1.7$\\
non-collision		& $5.33 \pm 1.95$ & $6.6 \pm 2.3$\\
$W/Z$ + jet             & $1.37 \pm 0.95$ & $3.1\pm 1.5$\\
$W\gamma$		& $0.90 \pm 0.14$ &$2.22\pm 0.3$\\ \hline
Total background	& $17.3 \pm 2.38$& $49.9 \pm 4.1$\\
$N_{\nu\bar\nu\gamma}^{\rm SM}$ & $33.7 \pm 3.4$&--\\\hline
$N_{\rm obs}$		& 51 & 51\\\hline
\end{tabular}
\end{center}
\end{table}

The signal process $\gamma G_{KK}$ is generated~\cite{priv:pythia}
using {\sc pythia} for number of extra dimensions
ranging from $2$ to $8$ and for a $M_{D}$ 
value of $1.5\ {\rm GeV/c^2}$.  A total luminosity-averaged 
efficiency of about
$43\%$ is estimated 
for central photons of $E_{T}>90\ {\rm GeV}$.
No NLO order K-factors are applied to the LED
signal.  The number of observed events ($N_{obs}$) can be
seen in Table~\ref{tab:bkg}.

In order to estimate the acceptance of 
$Z\gamma\to\nu\bar{\nu}\gamma$ 
events for
photons of  $E_{T}>90\ {\rm GeV}$, we use events from
MC samples produced
with a leading-order (LO) $Z\gamma$ 
generator~\cite{CT:baur_qcd:1998} that 
are passed through 
a parametrized simulation of the \D0 detector. 
The expected number of events 
($N^{SM}_{\nu\bar{\nu}\gamma}$) according
to this simulation and the number of observed 
events are shown in Table~\ref{tab:bkg}. 
The MC samples for the ATGC signal 
are produced with the LO $Z\gamma$
generator with values of $h_{30}^{V}$ and $h_{40}^{V}$
that differ from zero and for the form-factor scale
$\Lambda = 1.5\ {\rm TeV}$.

Next-to-leading order (NLO) QCD effects are 
predicted to be small for our event selection, 
and  we assign $7\%$ 
uncertainty to SM process cross sections due to them. Other sources of 
systematic uncertainty include uncertainty due to photon identification
($5\%$), choice of parton distribution functions ($7\%$) and uncertainty in
the total integrated luminosity ($6.1\%$).

\section{Results and Conclusions}

\begin{figure}
\begin{minipage}[t]{0.4\textwidth}
\includegraphics[scale=0.3]{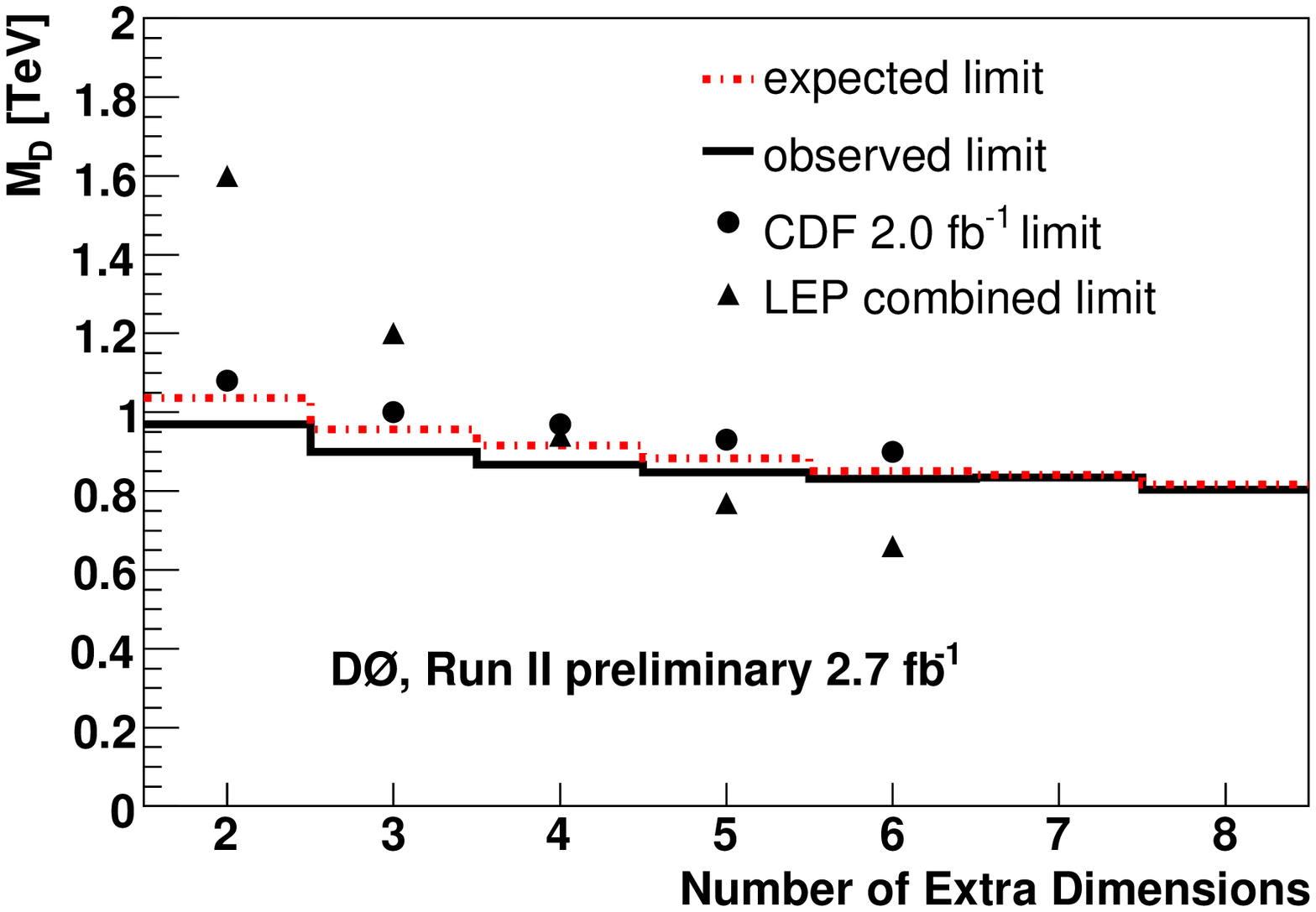}
\caption{\label{ledlimits} \small\small The
expected and observed lower limits on $M_{D}$ for
LED in the $\gamma + \met$ final state. 
CDF
limits with  $2\ {\rm fb}^{-1}$~of data (single photon
channel)~\cite{Aaltonen:2008hh}, and
the LEP combined limits~\cite{lepled} are also shown.}
\end{minipage} 
\hfill
\begin{minipage}[t]{0.55\textwidth}
\includegraphics[scale=0.2]{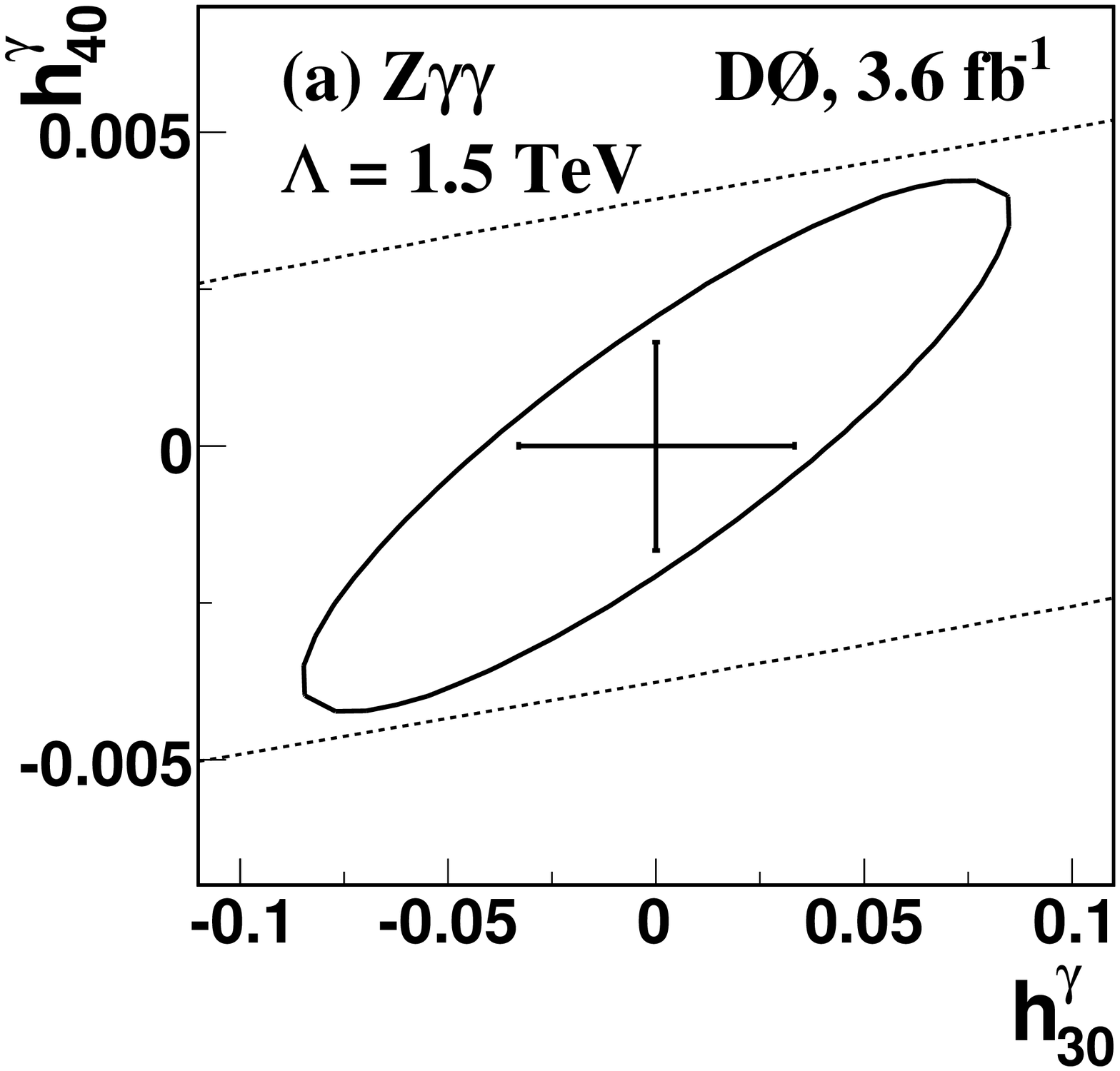}
\includegraphics[scale=0.2]{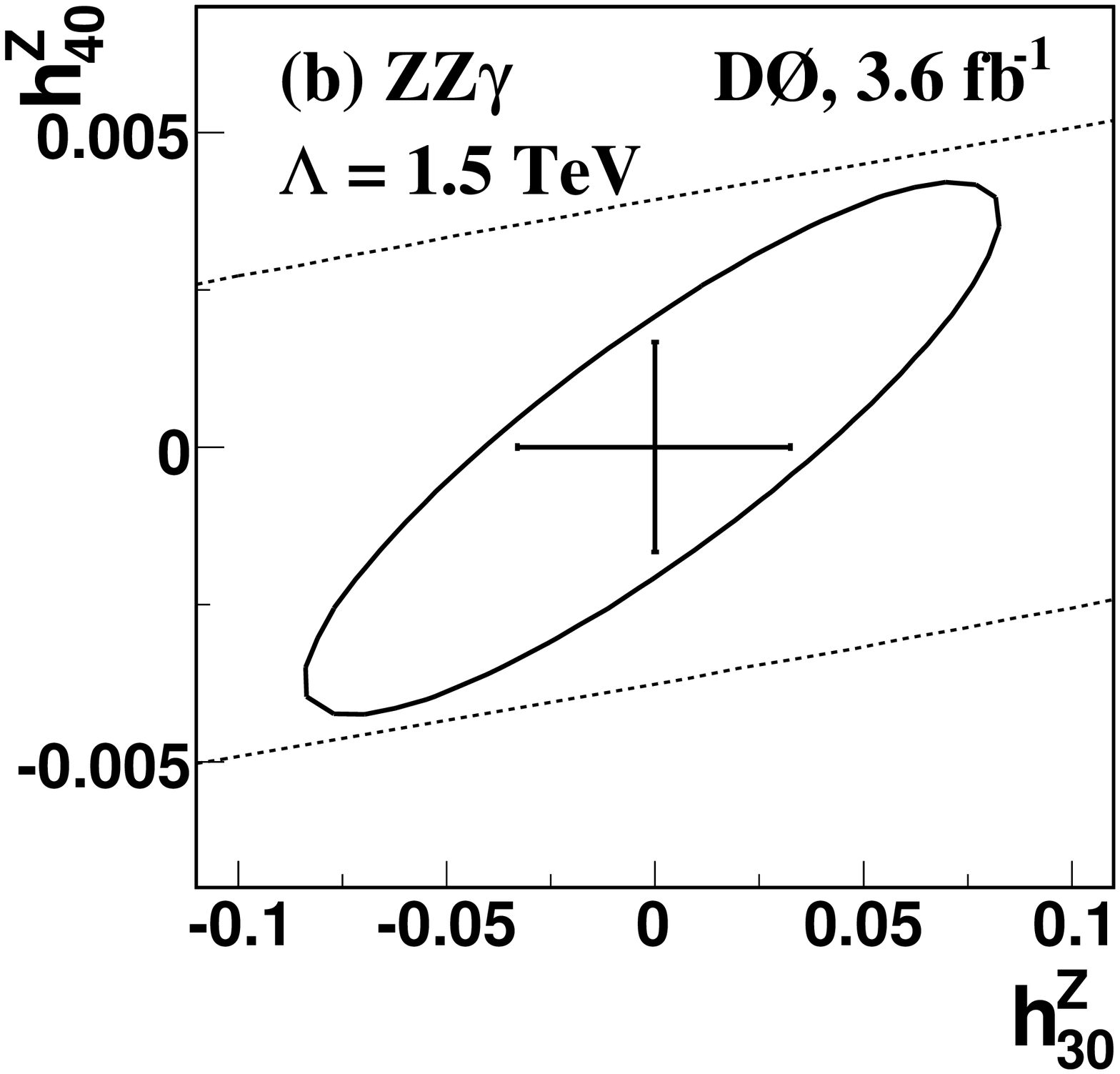}
\caption{\label{aclimits} \small Two-dimensional bounds (ellipses) 
at 95\% C.L. on CP-conserving (a) 
$Z\gamma\gamma$ and (b) $ZZ\gamma$ couplings. The crosses represent the one-dimensional bounds
at the 95\% C.L. setting all other couplings to zero. The dashed lines indicate 
the unitarity limits for $\Lambda = 1.5$~TeV.}
\end{minipage} 

\end{figure}

The number of single photon candidate events
are in agreement with the number of background events,
therefore we proceed to set limits.
In the case of LED analysis, we employ the modified frequentist 
approach~\cite{MAN:fisher,CT:junk}, 
using the binned photon $E_{T}$ distribution, 
to set limits for $M_{D}$ (Fig.~\ref{ledlimits}) at the $95\%$ confidence level (C.L.):
$M_{D}>970,899,867,848,831,834$ and
$804\ {\rm GeV/c^2}$ for $n=2,3,4,5,6,7$ and $8$ extra dimensions.



The $Z\gamma$ cross section multiplied by the BR 
of the invisible channel is measured to be
$32 \pm 9 {\rm (stat.+syst.)} \pm 2 {\rm (lumi.)~fb}$, which
in good agreement with the SM
NLO prediction of $39 \pm 4$~fb~\cite{CT:baur_qcd:1998}.  
The statistical
significance of the measured cross section
corresponds to $5.1$ standard deviations (s.d.).  Thus, this is
the first observation of the $Z\gamma\to\nu\bar{\nu}\gamma$ process
at the Tevatron.  Since the SM production and observed production
agree, we set limits on ATGC at the $95\%$ C.L. 
using a binned likelihood
method, which compares, bin by bin, the photon
$E_{T}$ spectrum in data 
with the corresponding distribution from
SM background and (``anomalous'') $Z\gamma$ signal.  The comparison
is performed for each pair of couplings in a two
dimensional grid involving  $h^{V}_{30}$ and $h^{V}_{40}$.
Poisson distributions is assumed for data and ATGC signal, while
uncertainties for the background, all systematic uncertainties and
luminosity are assumed to be Gaussian. 
To calculate one-dimensional limits we set the additional
anomalous coupling to zero. The one-dimensional limits in the
neutrino channel are:
$|h_{30}^{\gamma}| < 0.036$, $|h_{40}^{\gamma}| < 0.0019$
and $|h_{30}^{Z}| < 0.035$, $|h_{40}^{Z}| < 0.0019$.

We combine the results in the invisible decay channel of the 
$Z$ boson with the $Z\gamma \to \ell\ell\gamma$ ($\ell = e,~\mu$)
channels~\cite{zgplb}, 
which were studied in detail in~\cite{CT:p17_zg_plb} for 
1~fb$^{-1}$ of data.
The one-dimensional limits for the combination of the 
three channels are: 
$|h_{30}^{\gamma}| < 0.033$, $|h_{40}^{\gamma}| < 0.0017$
and $|h_{30}^{Z}| < 0.033$, $|h_{40}^{Z}| < 0.0017$.  
We can assert that the limits
on $h_{30}^{Z}$, $h_{40}^{Z}$, and $h_{40}^{\gamma}$
are the most restrictive to date.
One and two-dimensional limits can be appreciated in
Fig.~\ref{aclimits} 

To summarize, we tested the strength of the electroweak
force 
and searched for the presence of LED in our universe via
single photon plus missing energy final states
finding no hints of new physics.  We observed, for the first
time at the Tevatron, the $Z\gamma\to\nu\bar{\nu}\gamma$ process
at a statistical significance of $5.1$ s.d. and set
limits on the ATGC, some of which are the most restrictive
to date.

\section*{Acknowledgments}
The author wishes to thank 
the staffs at Fermilab and collaborating institutions, 
principally Alexey Ferapontov, Yuri Gershtein and Yurii
Maravin.

\section*{References}
\bibliography{iopart-num}

\begin{thebibliography}{10}
\bibitem{defeta}
The pseudo-rapidity $\eta$ is defined as $\eta = -ln[tan(\theta/2)]$, where
  $\theta$ is the polar angle measured with respect to the proton beam
  direction. $\phi$ is the azimuthal angle.
\bibitem{zgplb}
MC samples for these channels were regenerated with $\Lambda = 1.5$~TeV
\end{thebibliography}

\end{document}